      [1999/12/01 v1.4c Il Nuovo Cimento]
\documentclass{cimento}

\usepackage{amsfonts}
\usepackage{graphicx}

\title{On the MSSM with hierarchical squark masses and a heavier Higgs boson}
\author{E.~Bertuzzo\thanks{Review of \cite{Barbieri:2010pd}, to appear in the Proceedings of the LC10 workshop}}
\instlist{\inst{} Scuola Normale Superiore and INFN, Piazza dei Cavalieri 7, 56126 Pisa, Italy}
\begin{document}

\maketitle

\begin{abstract}
In the contest of supersymmetric extensions of the Standard Model, we consider a spectrum in which the lightest Higgs boson 
has mass between $200$ and $300$ GeV and the first two generations of squarks have masses above $20$ TeV, considering the Higgs 
boson mass and the Supersymmetric Flavour Problem as related naturalness problems. After the analysis of some models in which 
the previous spectrum can be naturally realized, we consider the phenomenological consequences for the LHC and for Dark Matter.
\end{abstract}

\section{Introduction and statement of the problem}
\begin{figure}[tb]
\begin{center}
\includegraphics[width=0.7\textwidth]{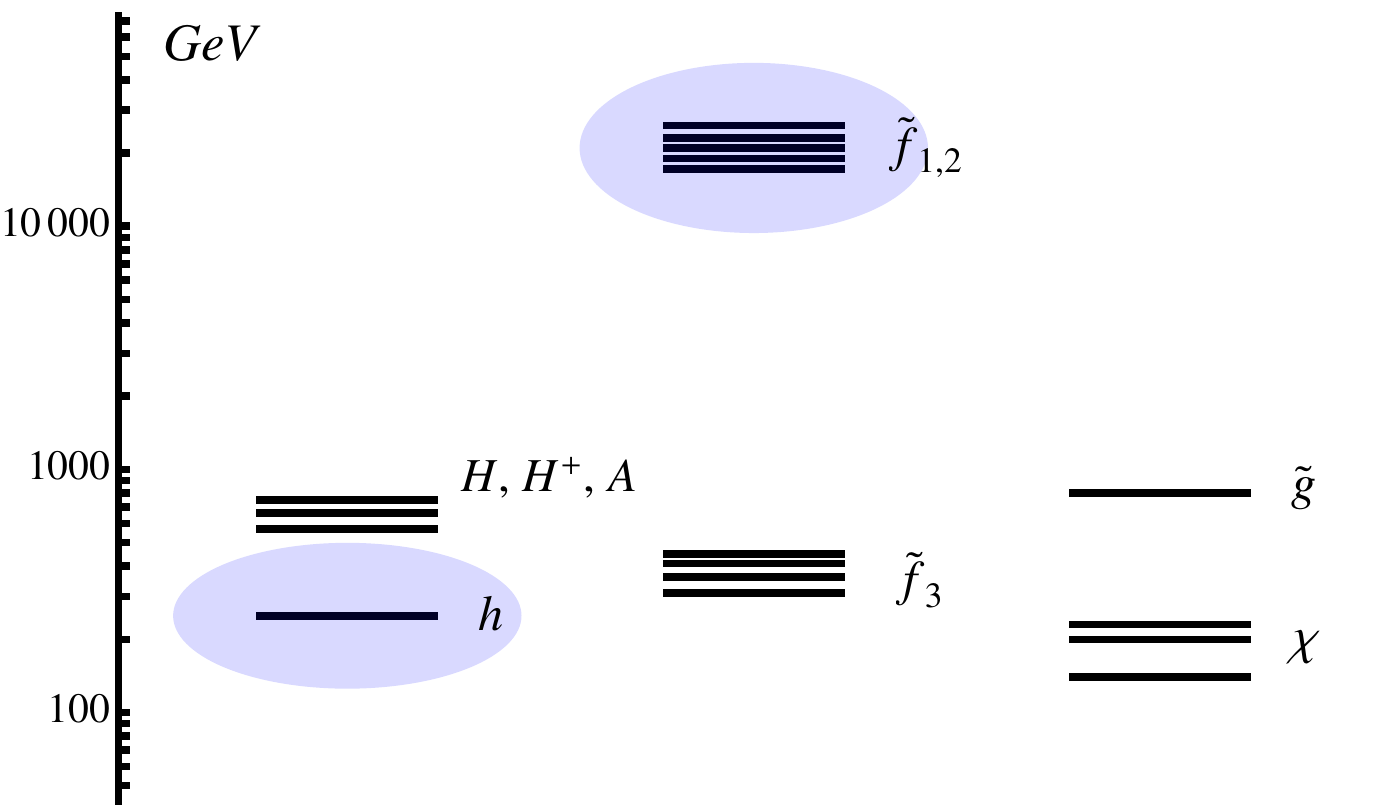}
\caption{A representative of the spectrum we are considering, with $m_h = (200\div 300)~\mathrm{GeV}$ and 
$m_{\tilde{q}_{1,2}}\geq 20~\mathrm{TeV}$.}
\label{fig:spettro} 
\end{center}
\end{figure}
\begin{figure}
\begin{center}
\includegraphics[width=0.6\textwidth]{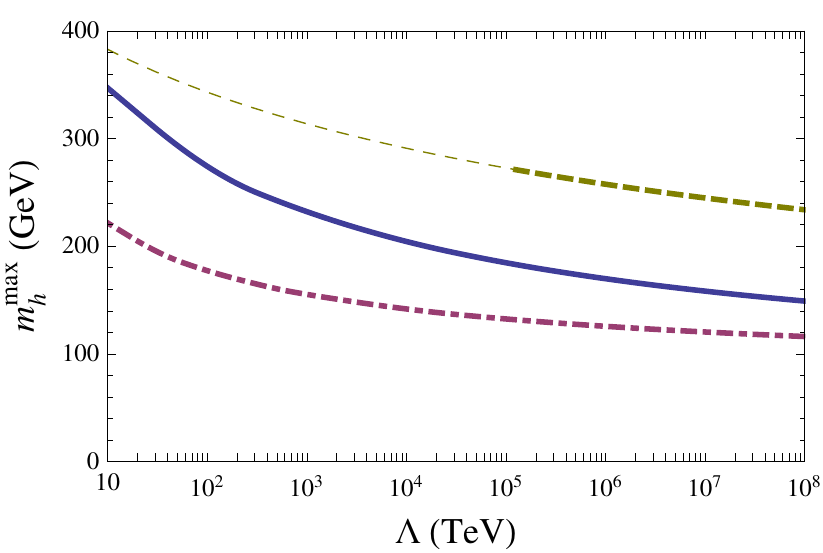}
\caption{Upper bounds on $m_h$ as a function of the scale $\Lambda$ where some couplings become semi-perturbative in the 
three different cases: extra $\mathrm{U(1)}$ (dotdashed), $\lambda$SUSY (solid) and $\mathrm{SU(2)}$ (dashed).}
\label{fig:max_mh} 
\end{center}
\end{figure}

Supersymmetry is surely one of the best motivated extensions of the Standard Model (SM). However, it is well known that 
the Minimal Supersymmetric Standard Model (MSSM) suffers for at least two phenomenological problems: on the one hand 
the MSSM predicts $m_h \leq m_Z |\cos 2\beta|$ as upper bound for the lightest Higgs boson mass at tree level, in potential 
conflict with the LEP II lower bound $m_h \geq 114$ GeV \cite{LEPHIGGS}. On the other hand, the MSSM general flavour structure predicts 
signals potentially in conflict with the present good agreement between the SM prediction and the data. As is well known, the first problem is a naturalness problem: the 
sensitivity of the Fermi scale on the average stop mass makes unnatural to raise the mass of the lightest Higgs boson much 
above the tree level upper bound \cite{Barbieri:1987fn}-\cite{Dimopoulos:1995mi}. At the same time, the flavour problem can be solved (or at least ameliorated) 
allowing the masses of the 
first two generations of squarks to be heavy enough to suppress unwanted signals \cite{Nir:1993mx}. Up to which value they can be 
pushed can be a naturalness problem 
as well, so that we argue in favour of a view in which the two issues, the ``Higgs problem'' and the ``Flavour problem'' 
may be related naturalness problems.\\
To be more precise, the more stringent bounds on the masses of the squarks of the first two generations come from the 
$\Delta S=2$ transitions, both real and especially imaginary \cite{Giudice:2008uk}; demanding only for heavy squark masses, one ends up with 
masses of the order of hundreds of TeV, while demanding also for degeneracy and alignment between the first two generations 
of order of the Cabibbo angle, the lower bounds are relaxed: \emph{i.e.} assuming $\delta^{LL} \gg \delta^{RR, LR}$ (where, according
 to the standard notation, $\delta \simeq \frac{|m_1^2-m_2^2|}{(m_1^2+m_2^2)/2}$ and LL, RR and LR refers to left and right 
sector, respectively) and $\delta^{LL} \simeq \lambda \simeq 0.22$ we have (for details, see \cite{Barbieri:2010pd})
\begin{equation}
\begin{array}{ccl}
 \Delta C=2 & \Rightarrow & m_{\tilde{q}_{1,2}} \geq 3~\mathrm{TeV}\\
 \mathrm{Im}(\Delta S)=2, \sin\varphi_{CP}\simeq 0.3 & \Rightarrow & m_{\tilde{q}_{1,2}} \geq 12~\mathrm{TeV}\\
\end{array}
\end{equation}
Let us now formulate in equations our starting point: the two naturalness bounds (where $1/\Delta$ is the amount of 
fine-tuning and $m_{\tilde{t}}$ is the average stop mass) \cite{Barbieri:1987fn}
\begin{equation}\label{eq:naturalness_bounds}
 \begin{array}{ccc}
  \frac{m^2_{\tilde{t}}}{m_h^2} \frac{\partial m_h^2}{\partial m^2_{\tilde{t}}}< \Delta & & 
\frac{m^2_{\tilde{q}_{1,2}}}{m_h^2} \frac{\partial m_h^2}{\partial m^2_{\tilde{q}_{1,2}}}< \Delta
 \end{array}
\end{equation}
must be considered together. It is clear from Eq. (\ref{eq:naturalness_bounds}) that increasing the Higgs boson mass goes 
in the direction of relaxing any naturalness bound, so that it is conceivable to have squarks 
of the first two generations with masses high enough to solve the flavour problem without introducing too much fine-tuning. In summary, we seek for models in which the spectrum of Fig. 
\ref{fig:spettro} 
can be realized in a natural manner.\\
Extensions of the MSSM that allow for a significant increase of the Higgs boson mass have been studied in the literature; a 
representative set is the following:
\begin{itemize}
 \item \emph{Extra $\mathrm{U(1)}$ factor} \cite{Batra:2003nj}. The MSSM gauge group is extended to include an additional $\mathrm{U(1)_X}$ factor 
with coupling $g_x$ and charge $\pm 1/2$ of the two standard Higgs doublet. The extra gauge factor is broken by two extra 
scalars, $\phi$ and $\phi_c$, at a scale significantly higher than $v$. The tree level upper bound on the mass of the 
lightest scalar becomes
\begin{equation}
 m^2_h \leq \left(m_Z^2 + \frac{g_x^2 v^2}{2\left( 1+\frac{M^2_X}{2 M^2_\phi}\right)}\right) \cos^2 2\beta
\end{equation}
where $M_X$ and $M_\phi$ are the masses of the gauge boson and the soft breaking mass of $\phi$ and $\phi_c$ taken approximately 
degenerate.
 \item \emph{Extra $\mathrm{SU(2)}$ factor} \cite{Batra:2003nj}-\cite{Batra:2004vc}. The extended gage group is now 
$\mathrm{SU(3)_c\times SU(2)_I\times SU(2)_{II}\times U(1)_Y}$ where the two $\mathrm{SU(2)}$ gauge groups are broken 
down to the diagonal subgroup by a chiral bidoublet $\Sigma$ at a scale much higher than the electroweak scale. The upper bound 
on the mass of the Higgs boson is now
\begin{equation}
m_h^2 \leq m_Z^2 \frac{g'^2+ \eta g^2}{g'^2+g^2}, ~~~~~\eta= \frac{1+ \frac{g^2_I M_\Sigma^2}{g^2 M_X^2}}
{1+\frac{M_\Sigma^2}{M_X^2}} 
\end{equation}
where $g_I$ is the gauge coupling associated to $\mathrm{SU(2)_I}$, $M_\Sigma$ the soft breaking mass of the $\Sigma$ scalar 
and $M_X$ the mass of the quasi degenerate heavy gauge triplet vectors.
 \item \emph{$\lambda$SUSY} \cite{Barbieri:2006bg}-\cite{Cavicchia:2007dp}. This is the NMSSM \cite{NMSSMreview} with a 
largish coupling $\lambda$ between the singlet and the two Higgs doublets. The upper bound in this case is
\begin{equation}
m_h^2 \leq m_Z^2 \left(\cos^2 2\beta + \frac{2 \lambda^2}{g^2 + g'^2}\sin^2 2\beta\right) 
\end{equation}
\end{itemize}
Fig. \ref{fig:max_mh} shows the maximal value of $m_h$ in the three different cases ($\tan\beta \gg 1$ in the extra-gauge cases 
and low $\tan\beta$ for $\lambda$SUSY) as a function of the scale at which the relevant coupling becomes semi-perturbative.
\section{Constraints from naturalness and from colour conservation}
\begin{figure}[tb]
\begin{center}
\begin{tabular}{cc}
\includegraphics[width=0.44\textwidth]{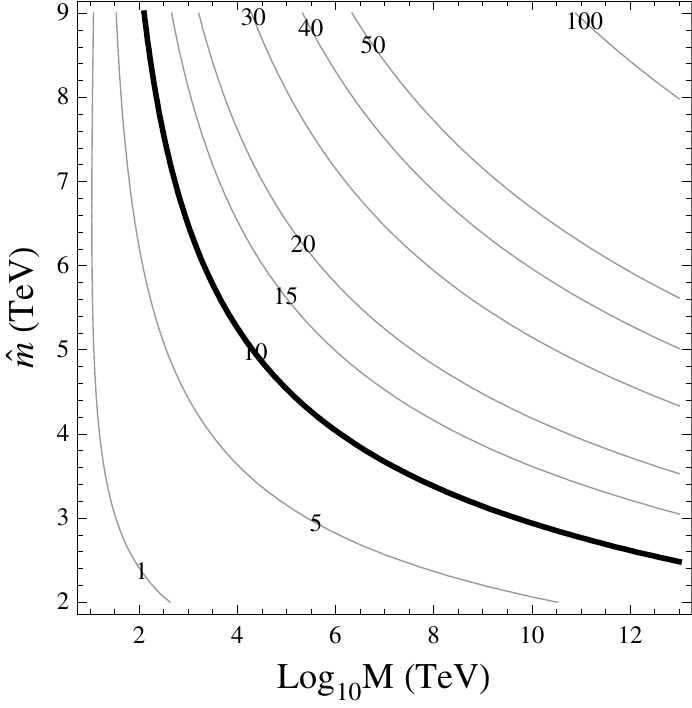} &
\includegraphics[width=0.46\textwidth]{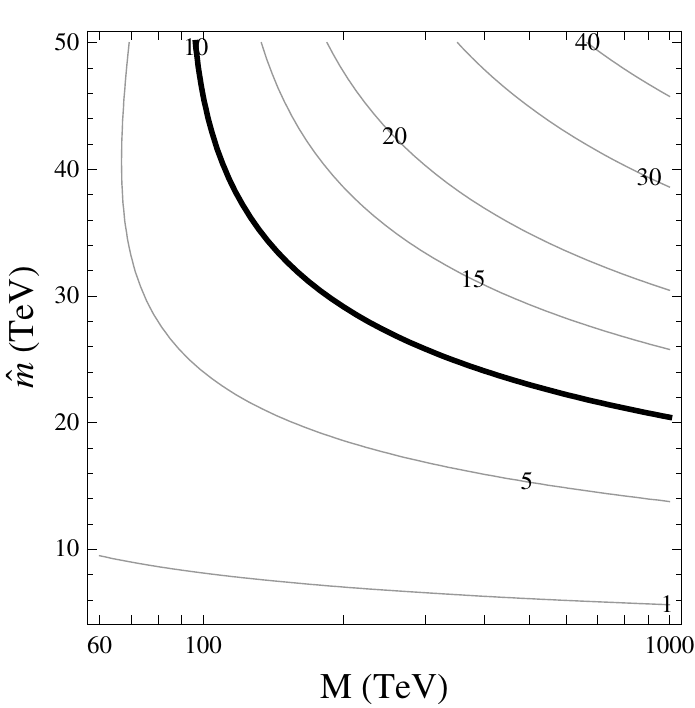}
\end{tabular}
\end{center}
\caption{Naturalness upper bounds on the common mass of the squarks of the first two generation as a function of the 
scale $M$ at which the renormalization group flow begins. Left panel: MSSM ($m_h=120~\mathrm{GeV}$, $\tan\beta\gg1$), Right 
panel: $\lambda$SUSY ($m_h=200~\mathrm{GeV}$, $\tan\beta\simeq 1$).}
\label{fig:naturalness_plot}
\end{figure}

\begin{figure}[tb]
\begin{center}
\begin{tabular}{cc}
\includegraphics[width=0.44\textwidth]{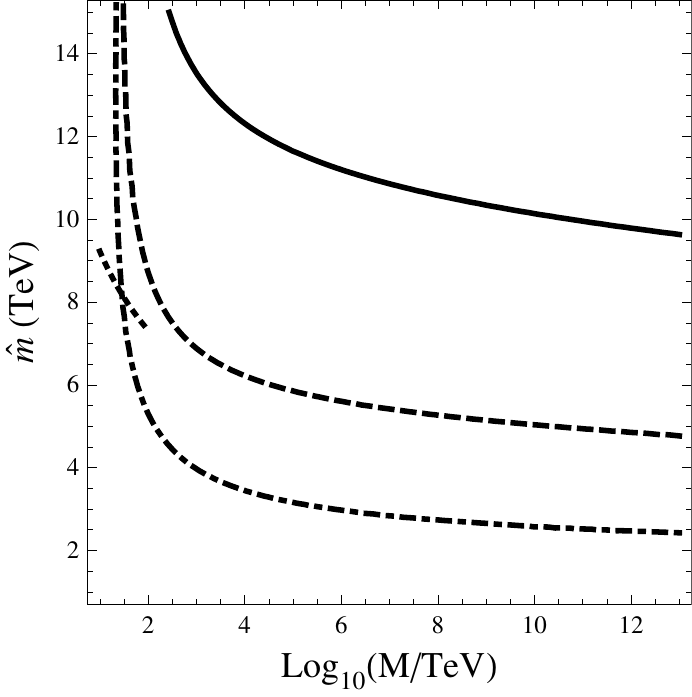} &
\includegraphics[width=0.46\textwidth]{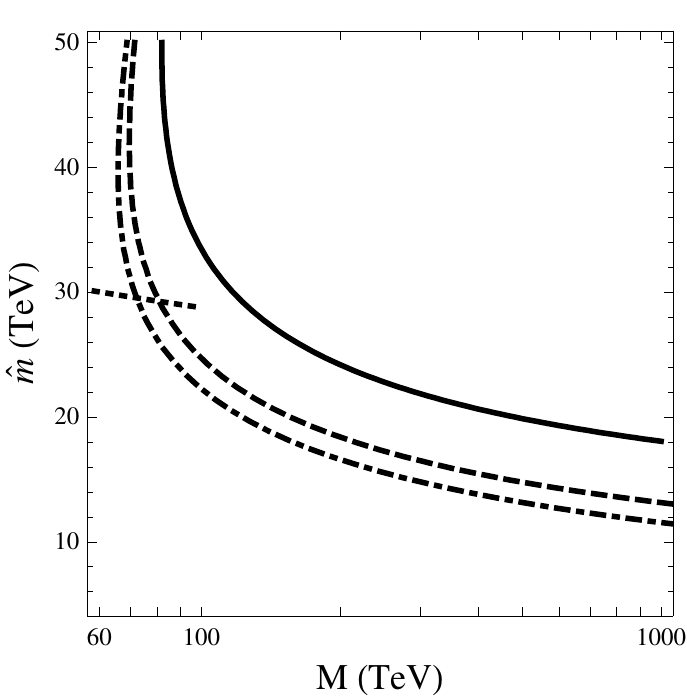}
\end{tabular}
\end{center}
\caption{Upper bounds on the common mass $\hat{m}$ coming from colour conservation for different values of the gluino mass: 
$2~\mathrm{TeV}$ (solid), $1~\mathrm{TeV}$ (dashed) and $0.5~\mathrm{TeV}$ (dotdashed). Left panel: MSSM ($m_h=m_Z$), Right 
panel: $\lambda$SUSY ($m_h=250~\mathrm{GeV}$).}
\label{fig:colour_plot}
\end{figure}

We can now discuss what happens to the naturalness bounds of Eqs.(\ref{eq:naturalness_bounds}) once we raise the Higgs boson 
mass. First of all, the bound on the stop mass is relaxed, but now its value is no longer relevant for the ``Higgs boson 
problem'', since the mass of the lightest Higgs boson is above the LEP bound already at tree level. Concerning the bound 
on the mass of the squarks of the first two generations, Fig. \ref{fig:naturalness_plot} shows the comparison between 
the MSSM ($m_h=115~\mathrm{GeV}$, $\tan\beta\gg1$) and $\lambda$SUSY ($m_h=250~\mathrm{GeV}$, $\tan\beta\simeq 1$), 
assuming a common mass $m_1=m_2=\hat{m}$ at the scale $M$ at which the RGE flow begins (for details, see 
\cite{Barbieri:2010pd}). We do not show the analogous plot for the two gauge extensions since the bounds are much stronger 
than the MSSM case.\\
A complementary issue we have to care about is colour conservation, since the large values of the masses of the squarks 
of the first two generations can drive to negative values the squared mass of the third one \cite{ArkaniHamed:1997ab}. To properly analyse the problem,
 we proceed as follows: first of all we take a value of $m_{\tilde{Q}_3}$ at $M$ that gives at most a 
$10\%$ fine-tuning on the Fermi scale; we then demand the running due to the squarks of the first two generation not to drive 
$m_{\tilde{Q}_3}$ to negative values. The upper bounds on $\hat{m}$ are shown in Fig. \ref{fig:colour_plot} for different 
values of the gluino mass, in the MSSM (left panel) and 
$\lambda$SUSY (right panel), for $m_h=m_Z$ and $m_h=250~\mathrm{GeV}$ respectively. As can be seen, they are 
similar or weaker than the corresponding bounds obtained from naturalness considerations.
\section{Phenomenology}
\begin{figure}
\begin{center}
\begin{tabular}{cc}
\includegraphics[width=0.44\textwidth]{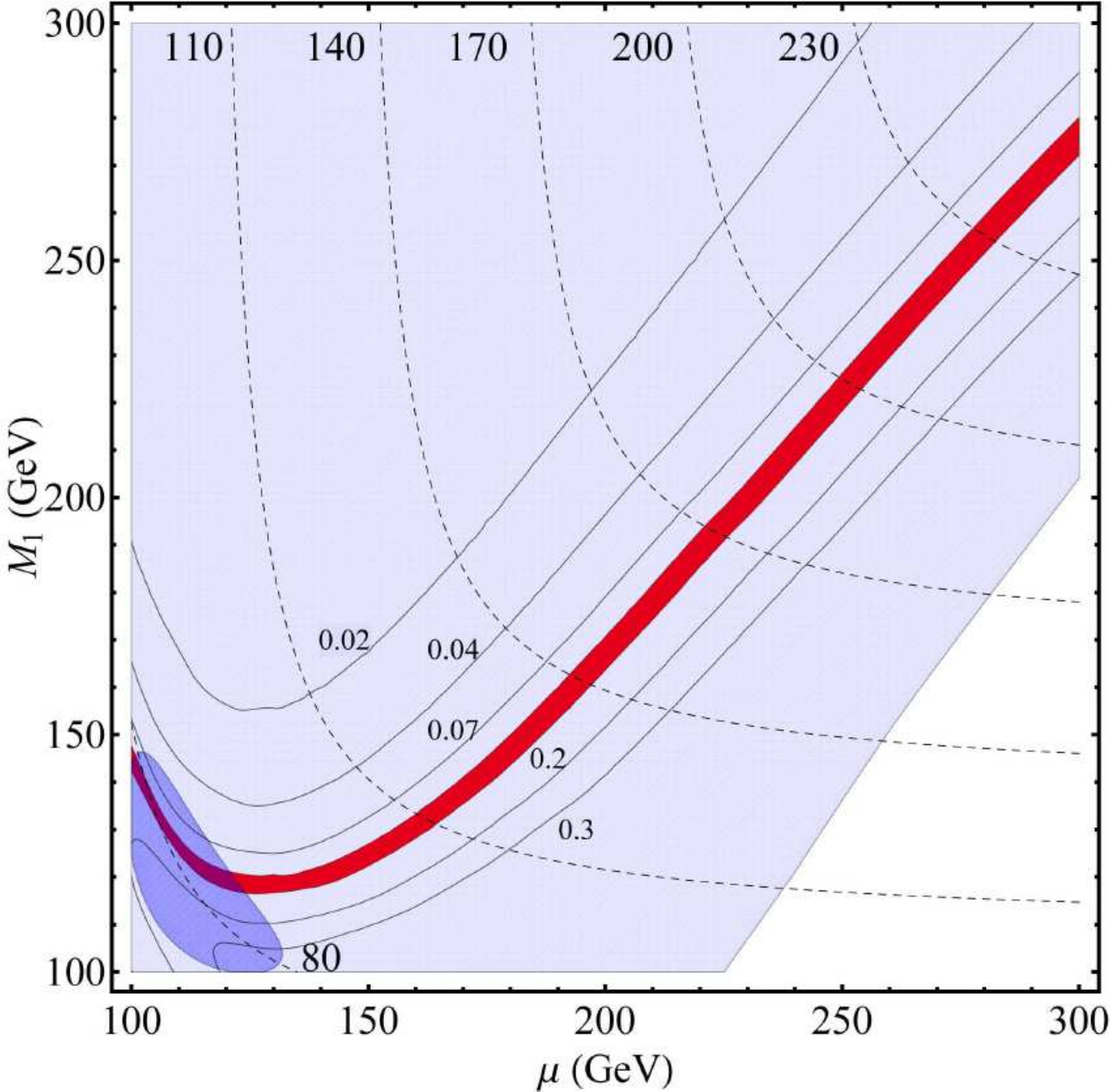} &
\includegraphics[width=0.44\textwidth]{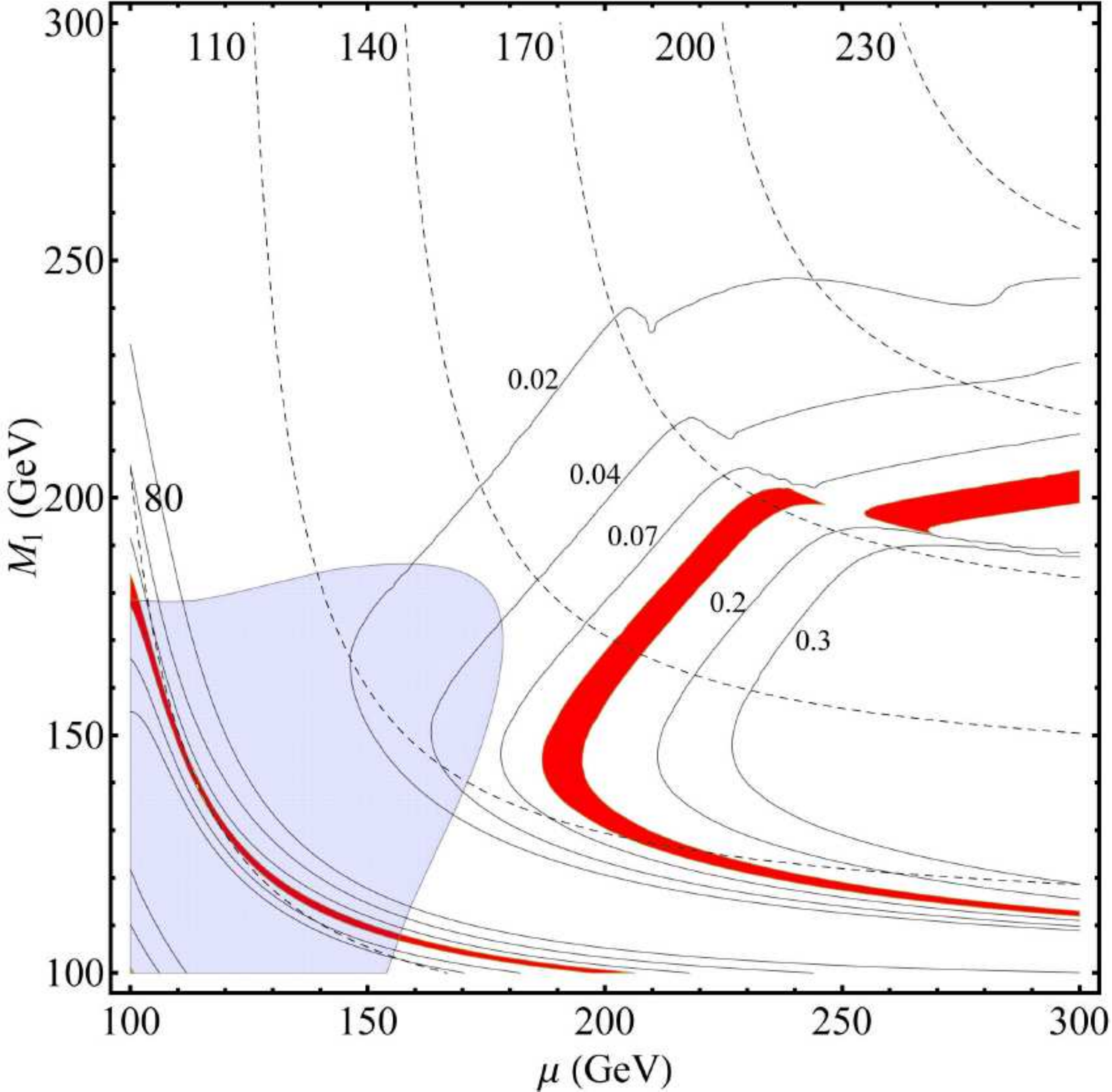}
\end{tabular}
\end{center}
\caption{Isolines of DM abundance (solid) and of LSP masses (dashed) for a decoupled Wino. Dark blue regions: CDMS exclusion, 
light blue: Xenon100 2010 exclusion projection. Left: MSSM, $m_h=120~\mathrm{GeV}$, $\tan\beta =7$. Right: $\lambda$SUSY, 
$m_h=200~\mathrm{GeV}$, $\tan\beta=2$.}
\label{fig:DM_plot}
\end{figure}

We now focus on the main phenomenological features of $\lambda$SUSY: sparticle production and 
decays at the LHC, Higgs boson phenomenology and Dark Matter Direct Detection.\\
\begin{itemize}
 \item It is well known that, at least in the first stage of the LHC run, the relatively more interesting 
signals will probably come from gluino pair production (at least for gluino masses not too large). An effective way to 
parametrize 
the signal is to consider the semi-inclusive Branching Ratios into $t\bar{t}\chi$ ($B_{tt}$), $t\bar{b}\chi$ ($B_{tb}$) and 
into $b\bar{b}\chi$ ($B_{bb}$) where $\chi$ stands for the LSP plus $W$ and/or $Z$ bosons. To an excellent approximation, 
\begin{equation}
 B_{tt}+ 2 B_{tb}+B_{bb} \simeq 1
\end{equation}
in most of the relevant parameter space \cite{Barbieri:2010pd},
 so that the final state of gluino pair production is $pp \rightarrow \tilde{g}\tilde{g} \rightarrow qq\bar{q}\bar{q}+\chi\chi$ 
with $q$ either a top or a bottom quark. A particularly interesting signal comes from same-sign dilepton production, with a 
Branching Ratio given by $BR(\ell^\pm \ell^\pm) = 2 B_\ell^2 \left(B_{tb}+B_{tt}\right)^2$ where $B_\ell = 21\%$ is the 
Branching Ratio of the $W$ boson into leptons. In a relevant portion of the parameter space, $BR(\ell^\pm \ell^\pm)= (2\div 4)\%$, unless 
the two sbottoms become the lightest squarks and/or $m_{\tilde{g}} \leq m_{LSP}+m_t$.
 \item As a consequence of the large Higgs boson mass, the most striking feature of $\lambda$SUSY is the discovery of the Golden Mode 
$h\rightarrow ZZ$ with two real $Z$ bosons. However, it must be stressed that such a signal depends on the chosen 
superpotential: indeed, in a non scale-invariant case \cite{lambdaSUSY}, the decoupling of the Singlet allows to simply have a 
heavier Higgs boson with standard couplings to fermions and gauge bosons, so that the Golden decay mode is typical. 
On the other hand, choosing a scale invariant superpotential \cite{Franceschini:2010qz}-\cite{Bertuzzo:2011ij}, in a relevant region of the 
parameter space 
the decay of the lightest Higgs boson into a pair of pseudoscalars is the dominant decay channel, so that the discovery 
potential relies essentially on the ability of analyse signatures coming from this decay. Also intermediate situations 
are conceivable \cite{Lodone:2011ax} in which, depending on the region of parameter space, both behaviours can be present.
 \item In $\lambda$SUSY the LSP can acquire a Singlino component, in contrast to what happens in the MSSM. Let us consider 
the case in which this component is negligible due to a decoupled Singlino, so that the LSP is as usual an 
higgsinos/gauginos admixture that must satisfy the ``Well-Temperament'' \cite{ArkaniHamed:2006mb} in order to reproduce correctly the Dark Matter (DM) 
relic abundance. To be more precise, let us focus on the well-tempered bino/higgsino with a decoupled wino. The situation 
is shown in Fig. \ref{fig:DM_plot} for the MSSM ($m_h=120~\mathrm{GeV}$, $\tan\beta=7$) and $\lambda$SUSY ($m_h=200~\mathrm{GeV}$, 
$\tan\beta=2$). The solid lines represent the DM abundance while the dashed lines are the LSP mass. The red region corresponds 
to a DM abundance compatible with the experiments, the dark blue region is the CDMS exclusion while the light blue region 
is the 2010 exclusion projection for Xenon100. As can be seen, in the MSSM case there is a precise correlation between 
$\mu$ and $M_1$, manifestation of the Well-Temperament. This is not the case for $\lambda$SUSY, in which the Well-Temperament 
is completely disrupted around the region corresponding to a resonant Higgs boson exchange in the s-channel. Moreover, the 
exclusion coming from the Direct Searches are much weaker in the $\lambda$SUSY case, since the spin-independent cross section 
of a DM particle on a nucleon falls off as $1/m_h^4$.
\end{itemize}
\section{Conclusions}
We considered the possibility of regarding the ``Higgs boson problem'' and the ``Supersymmetric Flavour Problem''
 as related naturalness problems, giving attention to models in which the Higgs boson mass is increased already at tree level. 
Among the considered possibilities, we found that in $\lambda$SUSY \cite{lambdaSUSY} an Higgs boson mass 
of $250\div 300~\mathrm{GeV}$ allows to raise the masses of the squarks of the first two generations up to $20~\mathrm{TeV}$ 
without introducing too much fine-tuning, softening in this way the Supersymmetric Flavour Problem. Among the main phenomenological consequences, it is interesting to stress the 
possibility of detecting the Golden Decay mode $h\rightarrow ZZ$ in association to typical Supersymmetric signals due 
to multi-top final states. Regarding the DM, and focusing on a bino/higgsino LSP, the effect of an increased Higgs boson mass is twofold: 
on the one hand the ``Well-Temperament'' pointed out in \cite{ArkaniHamed:2006mb} 
is completely disrupted; on the other hand, only a small portion of parameter space is constrained by direct detection 
experiments, since the cross section falls off as $1/m_h^4$.

\acknowledgments
This work is supported in part by the European Programme ``Unification in the LHC Era",  
contract PITN-GA-2009-237920 (UNILHC).

\end{document}